\documentclass[prl,aps,floats,twocolumn]{revtex4}

\usepackage[dvips]{graphicx}
\usepackage{amssymb} 
\usepackage{amsmath}
\usepackage{ctable}

\begin{document}

\title{$CPT$-Symmetric Universe}

\author{Latham Boyle$^1$, Kieran Finn$^{1,2}$ and Neil Turok$^1$} 

\affiliation{$^{1}$Perimeter Institute for Theoretical Physics, Waterloo, Ontario, Canada, N2L 2Y5 \\
$^{2}$School of Physics and Astronomy, University of Manchester, Manchester, UK M13 9PL}

\date{March 2018}
\begin{abstract}
We propose that the state of the universe does {\it not} spontaneously violate $CPT$.  Instead, the universe after the big bang is the $CPT$ image of the universe before it, both classically and quantum mechanically.  The pre- and post-bang epochs comprise a universe/anti-universe pair, emerging from nothing directly into a hot, radiation-dominated era.  $CPT$ symmetry selects a unique QFT vacuum state on such a spacetime, providing a new interpretation of the cosmological baryon asymmetry, as well as a remarkably economical explanation for the cosmological dark matter. Requiring only the standard three-generation model of particle physics (with right-handed neutrinos), a $\mathbb{Z}_2$ symmetry suffices to render one of the right-handed neutrinos stable. We calculate its abundance from first principles: matching the observed dark matter density requires its mass to be $4.8\times10^{8}~{\rm GeV}$.  Several other testable predictions follow: (i) the three light neutrinos are Majorana and allow neutrinoless double $\beta$ decay; (ii) the lightest neutrino is massless; and (iii) there are no primordial long-wavelength gravitational waves.  We mention connections to the strong CP problem and the arrow of time.
\end{abstract}

\maketitle

{\bf Introduction.} Observations reveal that, seconds after the Big Bang, the universe was described by a spatially-flat radiation-dominated FRW metric (plus tiny gaussian, adiabatic, purely-growing-mode scalar perturbations described by a nearly-scale-invariant power spectrum; and, so far, no primordial vector or tensor perturbations) \cite{Ade:2015xua}.  This is a clue about the origin of the universe, but what is it trying to tell us?  The conventional view is that, in order to explain these simple initial conditions, one should imagine that the radiation dominated era we see was preceded by an earlier hypothetical epoch of accelerated expansion called inflation.  

In this Letter (and a longer companion paper \cite{Boyle:2018rgh}) we take a different view.  Ignoring perturbations for the moment, the metric we see in our past is strikingly simple and analytic: $g_{\mu\nu}=a^{2}(\tau)\eta_{\mu\nu}$ where $\eta_{\mu\nu}$ is the flat Minkowski metric, and the scale factor $a(\tau)$ is just proportional to the conformal time $\tau$.  If we take this metric seriously, and follow $a(\tau)\propto\tau$ across the bang, we find that the analytically extended FRW background with $-\infty<\tau<\infty$ suddenly exhibits a new isometry: time reversal symmetry $\tau\to-\tau$.  It thus becomes possible to adopt the natural hypothesis that, contrary to naive appearances, the state of our universe does {\it not} spontaneously violate $CPT$.  

In this Letter we explore the hypothesis that $CPT$ is unbroken, explain how it provides novel explanations for a number of the observed features of our Universe, 
and point out some predictions that will be tested in the coming years. In particular, we find it yields a remarkably economical explanation of the cosmological dark matter: if the Universe is in its preferred $CPT$-symmetric vacuum, late-time observers see heavy sterile neutrinos emerging from the bang, for the same reason that distant observers see Hawking radiation emerging from a black hole.  In our opinion, this provides the most elegant and compelling dark matter model currently available.

{\bf Spacetime (Background and Perturbations).}  In this Letter we work at the level of QFT on curved space.  Before turning to the state of the QFT, in this section we first consider what $(C)PT$ says about the spacetime itself at a purely classical level \cite{Wald:1980nm}. Thus, we treat the metric and the radiation fluid using general relativity:

The line element may be written in standard ADM form: $ds^{2}=-N^{2}d\tau^{2}+h_{ij}(dx^{i}+N^{i}d\tau)(dx^{j}+N^{j}d\tau)$.   To describe our universe (a flat FRW spacetime plus small scalar, vector and tensor perturbations), we use ``comoving gauge" so that the $x^{i}={\rm const}$ threads are normal to the $\tau={\rm constant}$ slices ($N^{i}=0$), and the threads follow the stress-energy flow so that (for scalar perturbations) the momentum density $T^{i}_{0}$ also vanishes.  Then we can write the lapse as $N=a[1+\phi]$, and $h_{ij}$ as $h_{ij}=a^{2}[(1+2{\cal R})\delta_{ij}+2\gamma_{ij}^{(0)}+2\gamma_{ij}^{(1)}+2\gamma_{ij}^{(2)}]$, 
where $a=a(\tau)$ is the background scale factor, ${\cal R}$ is the ``comoving curvature perturbation," $\phi$ is another scalar perturbation related to ${\cal R}$ by the Einstein equation, and we have split the traceless perturbation $\gamma_{ij}$ into its scalar, vector and tensor parts: $\gamma_{ij}^{(0)}$, $\gamma_{ij}^{(1)}$, $\gamma_{ij}^{(2)}$ \cite{Kodama:1985bj, Mukhanov:1990me}.

Next, to treat spinors and $CPT$, just as we switch from the wave operator $\Box$ to its ``square root" (the Dirac operator $D\!\!\!\!/$\,\;\!), we switch from the line element $ds^{2}$ to its ``square root" (the tetrad $e^{a}$).   Thus, we write $ds^{2}=\eta_{ab}e^{a}e^{b}$, where $\eta_{ab}={\rm diag}\{-1,1,1,1\}$ is the Minkowski metric, and choose a local Lorentz gauge where the tetrad one forms $e_{a}=e_{a\mu}dx^{\mu}$ are $e_{0}=-N\,d\tau$ and $e_{i}=h_{ij}^{1/2}dx^{j}$.  The spacetime is $(C)PT$ symmetric in the sense that the tetrad geometry according to an observer who moves forward along the $x^{i}={\rm const}$ thread is identical to the tetrad geometry according to an observer who moves backward along the thread {\it and} reverses the spatial one forms $e^{i}\to-e^{i}$.  Equivalently, the tetrad at time $\tau$ after the bang is the reverse of the tetrad at the corresponding moment before the bang along the same thread: 
\begin{equation}
  \label{CPT_e}
  e_{\mu}^{a}(\tau,{\bf x})=-e_{\mu}^{a}(-\tau,{\bf x}).
\end{equation}
Let us unpack the implications of this simple constraint:

i) {\it Background geometry}: Eq.~(\ref{CPT_e}) implies that the scale factor is odd, $a(-\tau)=-a(\tau)$, with $a\propto\tau$ near the bang (as in the radiation era).  

If this picture is correct, so that the bang is a topologically enforced singularity, cosmological models in which $a(\tau)$ undergoes a nonsingular bounce at a minimum scale factor $a_{{\rm min}}>0$ are misguided.  

ii) {\it Scalar perturbations}: In fourier space, neglecting anisotropic stress, ${\cal R}$ satisfies ${\cal R}''+2(z'/z){\cal R}'+c_{s}^{2}k^{2}{\cal R}=0$, where $c_{s}^{2}=\delta p/\delta\rho$ is the sound speed, $k$ is the comoving wave number, $z^{2}\equiv a^{2}\epsilon$ and $\epsilon=(a'/a)'/(a'/a)^{2}-1$.  Near the bang, where $a\propto\tau$ and $c_{s}^{2}=1/3$, the general solution is ${\cal R}({\bf k},\tau)=\tau^{-1}[A({\bf k})\,{\rm sin}(c_{s} k\tau)+B({\bf k})\,{\rm cos}(c_{s} k\tau)]$.  The condition (\ref{CPT_e}) then sets $B({\bf k})=0$, eliminating the mode that is singular at the bang, and selecting the well-behaved mode that approaches a constant as $\tau\to0$.  

This is precisely the boundary condition responsible for producing the famous oscillations seen in the CMB power spectrum, with the correct phases.  {\it This observed phenomenon, usually attributed to inflation, is alternatively explained by a symmetry argument.}

Also note that density pertubations grow as we get further from the bang in either direction, and hence the thermodynamic arrow of time points {\it away} from the bang in both directions (to the future and past).  The possibility that the thermodynamic arrow of time might reverse is an old one (going back to the debates between Boltzmann and his contemporaries \cite{Boltzmann}), and has been invoked more recently in several interesting contexts \cite{Aguirre:2003ck, Carroll:2004pn, Barbour:2014bga, Barbour:2015sba, Barbour:2016qyy, Goldstein:2016cnb}.

iii) {\it Vector perturbations}: Neglecting anisotropic stress, the gauge-invariant vector metric perturbation $\sigma_{g}$ satisfies 
$\sigma_{g}'+2(a'/a)\sigma_{g}=0$ \cite{Kodama:1985bj}, so $\sigma_{g}({\bf k},\tau)=C({\bf k})/\tau^{2}$ near the bang.  In our chosen gauge,
$\sigma_{g}\propto\gamma_{ij}^{(1)}{}'$, so Eq.~(\ref{CPT_e}) implies that $\sigma_{g}$ (and hence the primordial vorticity, which is tied to $\sigma_{g}$ by the $0i$ Einstein equation) vanishes, again in agreement with observations.

iv) {\it Tensor perturbations}: neglecting anisotropic stress, $\gamma_{ij}^{(2)}$ satisfies $\gamma_{ij}^{(2)}{}''+2(a'/a)\gamma_{ij}^{(2)}{}'+k^{2}\gamma_{ij}^{(2)}\!=0$ so that $\gamma_{ij}^{(2)}({\bf k},\tau)=\tau^{-1}[A_{ij}({\bf k})\,{\rm sin}(k\tau)+B_{ij}({\bf k})\,{\rm cos}(k\tau)]$.  Now (\ref{CPT_e}) sets $B_{ij}({\bf k})=0$, eliminating the mode that is singular at the bang and selecting the mode that is well behaved.  

Note that, for each type of perturbation -- scalar, vector, and tensor -- the condition (\ref{CPT_e}) ``protects" the geometry near the bang by precisely eliminating the dangerous singular modes that would cause the breakdown of linear perturbation theory and destroy the smooth (Weyl) character of the singularity.  In this way of looking at it, the elimination of the singular modes is not a consequence of a boundary condition at $\tau=0$, but is instead enforced by the symmetry between past and future.

v) {\it $U\bar{U}$ pair}: Eq.~(\ref{CPT_e}) implies $e_{\mu}^{a}(0,{\bf x})=0$.  If we combine this with Stueckelberg's observation that an anti-particle is a particle whose worldline proper time runs counter to the time in the embedding spacetime \cite{Stueckelberg:1941, Stueckelberg:1941rg}, it becomes natural to reinterpret the contracting half of our universe as an anti-universe (whose intrinsic proper time runs counter to the natural time-like coordinate in the embedding superspace, {\it i.e.}, the scale factor), so that our $CPT$-invariant universe is reinterpreted as a universe/anti-universe pair ($U\bar{U}$), emerging from nothing!  This interpretation continues to be useful when spinors and anti-particles enter the story: {\it e.g.}, the matter/anti-matter asymmetry on one side of the bang is the opposite of the asymmetry on the other side~\cite{Boyle:2018rgh}.  To convert this suggestive, semi-classical picture into a fully quantum one (as Feynman did with Stueckelberg's idea) is an important task, beyond the scope of this Letter.

{\bf $CPT$ Invariant Vacuum.}  Now we turn from the spacetime to the state of the QFT living on it.  

In Minkowski spacetime, there is a unique vacuum that respects the Minkowski isometries (more precisely, spacetime translations, Lorentz transformations, and $CPT$).  But in a generic curved spacetime, the choice of vacuum is ambiguous: different observers will naturally define different, inequivalent vacua, so that the zero particle state according to one observer will contain particles according to a different observer \cite{BirrellDavies}.  In particular, in an ordinary FRW spacetime, the isometries (spatial translations, spatial rotations, and parity) are not enough to determine a preferred vacuum, and comoving observers at different epochs will disagree.  But, as we explain in this section, if the FRW background also has an isometry under time reversal $\tau\to-\tau$, then there is a preferred vacuum that respects the full isometry group (including $CPT$).

Consider a spinor $\Psi$ with mass $m>0$ on a flat FRW background $ds^{2}=a^{2}(\tau)[-d\tau^{2}+d{\bf x}^{2}]$.  Its Lagrangian is
\begin{subequations}
  \label{L_psi}
  \begin{eqnarray}
    L&=&\sqrt{-g}[i\bar{\Psi}e_{a}^{\mu}\gamma^{a}\nabla_{\mu}\Psi-m\bar{\Psi}\Psi] \\
    &=&i\bar{\psi}\partial\!\!\!\!\;\!/\;\!\psi-\mu\bar{\psi}\psi.
  \end{eqnarray}
\end{subequations}
In the first line, we have the usual curved space Dirac operator \cite{BirrellDavies}; in comoving/conformal  coordinates, the tetrad is $e_{a}^{\mu}=(1/a)\delta_{a}^{\mu}$, and $\gamma^{a}$ are the $4\times4$ Dirac gamma matrices.  In the second line, we introduce the Weyl invariant spinor field $\psi(\tau,{\bf x})$ and its effective mass $\mu(\tau)$:
\begin{equation}
 \psi \equiv a^{3/2}\Psi,\qquad\quad\mu\equiv a\, m,
\end{equation}
with $\partial\!\!\!/\equiv\gamma^{\mu}\partial_{\mu}$ the flat-space Dirac operator, and partial derivatives $\partial_{\mu}$ with respect to $x^\mu=\{\tau,{\bf x}\}$.  The resulting equation of motion is
\begin{equation}
  \label{Dirac_eq}
  (i\partial\!\!\!\!\;\!/\,-\mu)\psi=0.
\end{equation}
Note that, since $a(\tau)$ is an odd function of $\tau$, so is $\mu(\tau)$.

To quantize, we expand $\psi(x)$ in a basis of solutions of Eq.~(\ref{Dirac_eq}), $\psi({\bf k},h,x)$ and $\psi^{c}({\bf k},h,x)$: 
\begin{equation}
  \label{psi_expansion}
  \sum_{h}\!\int\!\frac{d^{3}{\bf k}}{\!(2\pi)^{3/2}}[a({\bf k},h)\psi({\bf k},h,x)+b^{\dagger}({\bf k},h)\psi^{c}({\bf k},h,x)].
\end{equation}
Here $\psi({\bf k},h,x)\propto{\bf e}^{i{\bf k}{\bf x}}$ is the solution with momentum ${\bf k}$, helicity $h$, and ``positive frequency"; $\psi^{c}({\bf k},h,x)\equiv-i\gamma^{2}\psi^{\ast}({\bf k},h,x)$ is the charge-conjugate (``negative frequency") solution; and $a({\bf k},h)$ and $b({\bf k},h)$ are particle and anti-particle annihilation operators, which satisfy the usual fermionic anti-commutation relations: $\{a({\bf k},h),a^{\dagger}({\bf k}',h')\}\!=\!\{b({\bf k},h),b^{\dagger}({\bf k}',h')\}\!=\!\delta({\bf k}-{\bf k}')\delta_{h,h'}$, all other anti-commutators vanishing.

But in a general curved spacetime, there is no canonical choice for which solutions have ``positive frequency," and observers in different regions will make inequivalent choices: {\it e.g.}\ in FRW the positive frequency solutions $\psi_{-}$ and $\psi_{+}$ chosen, respectively, by observers in the far past ($\tau\to-\infty$) or far future ($\tau\to+\infty$) exhibit positive frequency behavior in these two limits, respectively:
$\psi_{\pm}({\bf k},h,x)\sim{\rm exp}[-i\!\int^{\tau}\!\!\!\omega(k,\tau')d\tau']$ as $\tau\to\pm\infty$, where $k=|{\bf k}|$ is the comoving wave number, and $\omega=\sqrt{k^{2}+\mu^{2}}>0$ is the comoving frequency.  The ``$-$" solutions may then be expressed in the ``$+$" basis:
\begin{equation}
  \label{psim_from_psip}
  \psi_{-}({\bf k},h,x)=\alpha({\bf k})\psi_{+}({\bf k},h,x)+\beta({\bf k})\psi_{+}^{c}(-{\bf k},h,x).
\end{equation}
We may adjust the phases of $\psi_{+}$ and $\psi_{-}$ so that $\alpha({\bf k})={\rm cos}\;\!\lambda({\bf k})$, $\beta({\bf k})=i\;\!{\rm sin}\;\!\lambda({\bf k})$, and $\lambda(-{\bf k})=-\lambda({\bf k})$ is real.  The ``$-$" observer's annihilation operators $(a_{-},b_{-})$ are then related to the ``$+$" observer's annihilation operators $(a_{+},b_{+})$ by the Bogoliubov transformation
\begin{equation}
  \left[\!\begin{array}{c} a_{+}(+{\bf k},h) \\ b_{+}^{\dagger}(-{\bf k},h) \end{array}\!\right]\!=
  \!\left[\!\begin{array}{rr} {\rm cos}\;\!\lambda({\bf k}) & i\;\!{\rm sin}\;\!\lambda({\bf k}) \\
  i\;\!{\rm sin}\;\!\lambda({\bf k}) & {\rm cos}\;\!\lambda({\bf k}) \end{array}\!\right]\!
  \!\left[\!\begin{array}{c} a_{-}(+{\bf k},h) \\ b_{-}^{\dagger}(-{\bf k},h) \end{array}\!\right]\!.
\end{equation}
The observer in the far past (resp. far future) defines the vacuum to be the state $|0_{-}\rangle$ (resp. $|0_{+}\rangle$) that is annihilated by all the operators $a_{-}$ and $b_{-}$ (resp. $a_{+}$ and $b_{+}$): $a_{\pm}({\bf k},h)|0_{\pm}\rangle=b_{\pm}({\bf k},h)|0_{\pm}\rangle=0$.  We are in the Heisenberg picture, so states do not evolve.  Note that, unless ${\rm sin}\;\!\lambda({\bf k})$ is identically zero, $|0_{-}\rangle$ and $|0_{+}\rangle$ are inequivalent: $|0_{-}\rangle$ has no particles according to its {\it own} particle number operator $N_{-}=a_{-}^{\dagger}a_{-}^{}$, but a non-zero number according to $N_{+}=a_{+}^{\dagger}a_{+}^{}$.  Moreover, since $a$ and $b$ transform as $[CPT]a_{\pm}({\bf k},h)[CPT]^{-1}=-b_{\mp}({\bf k},-h)$ and $[CPT]b_{\pm}({\bf k},h)[CPT]^{-1}=-a_{\mp}({\bf k},-h)$, the inequivalent vacua $|0_{+}\rangle$ and $|0_{-}\rangle$ are exchanged by $CPT$, $|0_{\pm}\rangle=CPT|0_{\mp}\rangle$, so neither vacuum is $CPT$ invariant.  

However, if we define new operators $a_{0}$ and $b_{0}$:
\begin{equation}
  \renewcommand{\arraystretch}{1.4}
  \left[\!\!\begin{array}{c} a_{0}(+{\bf k},h) \\ b_{0}^{\dagger}(-{\bf k},h) \end{array}\!\!\right]\!\!=\!
  \!\left[\!\!\begin{array}{rr} {\rm cos}\frac{\lambda({\bf k})}{2} & \;\mp i\;\!{\rm sin}\frac{\lambda({\bf k})}{2} \\
  \mp i\;\!{\rm sin}\frac{\lambda({\bf k})}{2} & {\rm cos}\frac{\lambda({\bf k})}{2} \end{array}\!\!\right]\!\!
  \!\left[\!\!\begin{array}{c} a_{\pm}(+{\bf k},h) \\ b_{\pm}^{\dagger}(-{\bf k},h) \end{array}\!\!\right]\!,\!
\end{equation}
they transform as $[CPT]a_{0}({\bf k},h)[CPT]^{-1}=-b_{0}({\bf k},-h)$ and $[CPT]b_{0}({\bf k},h)[CPT]^{-1}=-a_{0}({\bf k},-h)$, so the corresponding vacuum defined by $a_{0}|0_{0}\rangle=b_{0}|0_{0}\rangle=0$ is $CPT$ invariant: $CPT|0_{0}\rangle=|0_{0}\rangle$.  
In fact, there is a continuous family of $CPT$-invariant vacua, obtained by  defining $(a_{\eta}(+{\bf k},h), b_{\eta}^{\dagger}(-{\bf k},h))$ to be a real $SO(2)$ rotation of $(a_{0}(+{\bf k},h), b_{0}^{\dagger}(-{\bf k},h))$ through an angle $\eta$ satisfying $\eta({\bf k})=-\eta(-{\bf k})$. The vacuum defined by $a_{\eta}|0_{\eta}\rangle = b_{\eta}|0_{\eta}\rangle=0$ is still invariant under the full isometry group of the FRW background including $CPT$. However, among this family of ``$\eta$ vacua" the vacuum $|0_{0}\rangle$ is preferred since it minimizes the Hamiltonian in the asymptotic ``$+/-$" regions (or the particle number according to an early or late time observer)~\cite{Boyle:2018rgh}.

Now we assume the universe is in the preferred $CPT$-invariant vacuum state and consider the consequences:

{\bf Neutrino Dark Matter.}  Consider the standard model of particle physics (including a right-handed neutrino in each generation) coupled to Einstein gravity.  There is only one possible dark matter candidate in this model -- one particle that has not yet been detected and can have a lifetime longer than the age of the universe -- namely, one of the three right-handed neutrinos $\nu_{R,1}$.  This particle appears in two places in the Lagrangian: the Majorana mass term $\bar{\nu}_{R,i}^{c}M_{ij} \nu_{R,j}$ (where $M_{ij}$ is the $3\times 3$ Majorana mass matrix) and the Yukawa term $\bar{l}_{L,i}Y_{ij}\nu_{R,j}h_{c}$ (where $l_{L,j}$ is the left-handed lepton doublet, $h_{c}=i\sigma^{2}h^{\ast}$ is the charge conjugate of the Higgs doublet $h$, and $Y_{ij}$ is a $3\times3$ Yukawa coupling matrix).  The assertion that $\nu_{R,1}$ is exactly stable corresponds to the statement that the standard model couplings respect the $\mathbb{Z}_{2}$ symmetry $\nu_{R,1}\to-\nu_{R,1}$.  This symmetry sets to zero the first column of the matrix $Y_{ij}$, whose three entries $Y_{i1}$ would otherwise lead the $\nu_{R,1}$ to decay. 

Thus, in the same limit that $\nu_{R,1}$ becomes stable, it also becomes decoupled from all of the other particles in the standard model, and so might seem to become a poor dark matter candidate since it is not produced by thermal processes in the early universe.  But, in our picture, these particles have a predictable non-zero cosmic abundance, according to late-time comoving observers like us, just because the universe is in the $CPT$-invariant vacuum $|0_{0}\rangle$, which differs from our late-time vacuum $|0_{+}\rangle$.  If the stable neutrino's mass has a certain value, it automatically has the abundance, coldness and darkness needed to match observations.  This yields a strikingly simple alternative explanation for the dark matter, different from previous neutrino dark-matter models based on thermal or resonant production mechanisms \cite{Dodelson:1993je, Shi:1998km, Abazajian:2006yn, Boyarsky:2009ix, Canetti:2012kh}.

To see this explicitly, note that near the bang, {\it i.e.}, during the radiation era, above the electroweak phase transition, when $a(\tau)\propto\tau$,
the dark matter neutrino has equation of motion:
\begin{equation}
  \label{Dirac_eq_N}
  (i\partial\!\!\!\!\;\!/\,-\mu)N_{1}=0\qquad(\mu=\gamma\tau).
\end{equation}
Here $N_{1}\equiv a^{3/2}(\nu_{R,1}^{}+\nu_{R,1}^{c})$ and $\gamma$ is a constant given by $\gamma=(m_{dm}/m_{pl})\sqrt{\rho_{1}}$, where $m_{dm}$ is the mass of the right-handed neutrino $\nu_{R,1}$, $m_{pl}=(8\pi G/3)^{-1/2}\approx 4\times10^{18}~{\rm GeV}$ is the Planck scale, and $\rho_{1}=a^{4}\rho$ (the radiation density times $a^{4}$) is a constant.  

To understand the behavior of Eq.~(\ref{Dirac_eq_N}), consider the comoving frequency $\omega(\tau)=\sqrt{k^{2}+\mu^{2}}$.  If $\omega$ were independent of $\tau$, the solutions would be $N_{1}\propto {\rm e}^{i({\bf k}{\bf x}-\omega\tau)}$, just as in Minkowski space.  But since $\omega$ {\it does} depend on $\tau$, we turn to the WKB method.  Consider the dimensionless WKB parameter $|\omega'(\tau)/\omega^{2}|$: for fixed comoving wave number $k$, this vanishes near the bang (as $|\tau|\to0$) and far from the bang (as $|\tau|\to\infty$), but reaches a maximum value $\sim\gamma/k^{2}$ at an intermediate conformal time $|\tau_{max}|\sim k/\gamma$.   Thus, for wave numbers $k\gg \gamma^{1/2}$, the WKB parameter is always $\ll1$, WKB remains good and the Bogoliubov transformation between $(a_{-},b_{-})$ and $(a_{+},b_{+})$ is trivial: ${\rm sin}\;\!\lambda({\bf k})\approx 0$.   On the other hand, for wave numbers $k\ll \gamma^{1/2}$, WKB is badly violated, and the Bogoliubov transformation is maximal: $|{\rm sin}\;\!\lambda({\bf k})|\approx1$.  To see this, consider $\omega=\sqrt{\mu^{2}+k^{2}}$: in the limit $k\ll\gamma^{1/2}$, $\mu^{2}$ dominates and $k^{2}$ is negligible unless the mode is far outside the Hubble horizon; so we can neglect the spatial gradient terms in (\ref{Dirac_eq_N}) and solve $(i\gamma^{0}\partial_{\tau}-\mu)\psi=0$ to find $\psi=\hat{\psi}\,{\rm exp}[i{\bf k}{\bf x}+i\int^{\tau}\mu(\tau')d\tau']$ where $\hat{\psi}=(\xi,-\xi)$, with $\xi$ a constant 2-spinor.  Since $\mu(\tau)$ is odd, the solution switches from purely positive frequency in the far past to purely negative frequency in the far future (corresponding to $|{\rm sin}\;\!\lambda({\bf k})|=1$).  

Thus, for late time observers like us, the number density $n_{dm}$ of dark matter particles is $n_{dm}=(2\pi a)^{-3}\sum_{h}\int d^{3}{\bf k}\langle 0_{0}|N_{+}|0_{0}\rangle$ where the matrix element is $|{\rm sin}\,(\lambda({\bf k})/2)|^{2}$, so that $n_{dm}\sim(2\pi a)^{-3}\gamma^{3/2}\sim(2\pi)^{-3}(m_{dm}/m_{pl})^{3/2}\rho^{3/4}$, where $\rho^{3/4}\sim s$, the radiation entropy density.  Since the ratio $n_{dm}/s$ is conserved during the subsequent expansion, it must match the present day value $n_{dm,0}^{}/s_{0}$, where $s_{0}\sim2.3\times10^{-38}{\rm GeV}^{3}$ \cite{KolbTurner}, $n_{dm,0}=\rho_{dm,0}/m_{dm}$ is the present dark matter number density, and $\rho_{dm,0}\sim9.7\times10^{-48}~{\rm GeV}^{4}$ is the present dark matter energy density \cite{Ade:2015xua}.  Thus, we estimate $m_{dm}\sim[(\rho_{dm,0}/s_{0})(2\pi)^{3}m_{pl}^{3/2}]^{2/5}\approx {\rm few}\times10^{8}~{\rm GeV}$.  A more precise calculation \cite{Boyle:2018rgh} yields $m_{dm}=4.8\times10^{8}~{\rm GeV}$.

We emphasize that the definition of $|0_{0}\rangle$, and the resulting estimate of $n_{dm}$, is controlled by $CPT$ symmetry, not by the detailed physics of the bang itself.  In particular, we have seen that the Bogoliubov transformation is insensitive to the behavior of $a$ (or $\mu$) near $\tau=0$, where the WKB parameter vanishes, and is instead dominated by the WKB bump experienced by modes of wave number $k\sim\gamma^{1/2}$ at a proper time $t\sim m_{dm}^{-1}$ before or after the bang (when the temperature is already orders of magnitude below the Planck scale, and the usual radiation-dominated Friedmann equation should be reliable).  

{\bf Other Predictions.}  Several other predictions follow \cite{Boyle:2018rgh}:  (i) The three light neutrino mass eigenstates are Majorana particles (which will be tested by future neutrinoless double $\beta$-decay searches \cite{GomezCadenas:2011it}), and one of them is exactly massless (which will be tested by future cosmological constraints on the sum of the light neutrino masses \cite{Abazajian:2011dt}).  (ii) We have focused on the stable right-handed neutrino, but the other two (unstable) right-handed neutrinos are thermally coupled and can explain the observed matter/anti-matter asymmetry by thermal leptogenesis~\cite{Fukugita:1986hr, Buchmuller:2005eh}.  (iii) Since gravitational waves are massless, the corresponding ``$+$" and ``$-$" vacua agree.  Thus, no long wavelength gravitational waves are produced by our mechanism. 

{\bf Discussion.}  Let us end with a few remarks:

i) Here we assumed a flat, radiation dominated FRW background.  In a forthcoming paper, we explain how this background arises \cite{BoyleTurok}.

ii) In this Letter, we have described the background spacetime geometry and radiation fluid purely classically, according to general relativity. A fuller treatment of the singularity to include the trace anomaly \cite{BirrellDavies} and quantum back-reaction requires semiclassical methods, involving complex classical solutions along the lines of~\cite{Gielen:2015uaa, Gielen:2016fdb, Feldbrugge:2017mbc}.   

iii) A fascinating open question is whether current observations allow the standard model or, more properly, its minimal extension incorporating neutrino masses, to remain valid all the way up to the Planck scale, or whether new physics is required below this scale.  With the measured central values of the Higgs and top quark masses, the Higgs quartic self-coupling $\lambda$ runs to negative values at an energy scale below the Planck mass \cite{Degrassi:2012ry, Buttazzo:2013uya}; however a recent analysis suggests that a strictly positive $\lambda$ all the way is only disfavored at the 1.5 or 2 $\sigma$ level \cite{Bednyakov:2015sca}.  Even if the Higgs effective potential runs negative at large vev,  finite temperature corrections are sufficient to stabilize the Higgs field at zero vev in the very early universe. There would only be an instability (to a negative-Higgs-potential bulk phase) at {\it late} cosmological times, far to our future. We find it intriguing that the most economical possibility, of no new physics, may be viable  \cite{Shaposhnikov:2009pv}, and might even explain the dark matter. 

iv) We have seen that stability of the dark matter neutrino $\nu_{R,1}$ implies that the Lagrangian has a symmetry under $\nu_{R,1}\to-\nu_{R,1}$.  This symmetry suffers from no anomalies -- not even gravitational anomalies~\cite{Witten}. It is well known that in the standard model, the lepton representations $\{l_{L},\nu_{R},e_{R}\}$ echo the quark representations $\{q_{L},u_{R},d_{R}\}$.  (This observation underlies Pati-Salam grand unification \cite{Pati:1974yy}, in which the leptons are a fourth color.)  The parallel symmetry in the quark sector, $u_{R,1}\to-u_{R,1}$,  is interesting for other reasons. Naively, it forces the bare mass of the up quark to zero which, in turn, solves the strong $CP$ problem \cite{Kaplan:1986ru}.  Unlike the symmetry we are using, this $\mathbb{Z}_{2}$ symmetry is anomalous due to the strong interactions; however, if it holds at {\it any} energy and, in particular, at a very high energy scale, this may be sufficient to solve the strong $CP$ problem~\cite{Srednicki:2005wc}.  A deeper understanding of these symmetries will likely require new insights into the origin of the three generations in the standard model.

{\bf Note added.}
Shortly after our Letter appeared on the arXiv, a follow-up paper \cite{Anchordoqui:2018ucj} pointed out that the ANITA experiment may have already seen evidence for our dark matter candidate.  

{\bf Acknowledgements.} We thank Claudio Bunster, Job Feldbrugge, Angelika Fertig, Steffen Gielen, Jaume Gomis, David B.~Kaplan, Ue-Li Pen, Laura Sberna and Edward Witten for discussions.  Research at Perimeter Institute is supported by the Government of Canada through Innovation, Science and Economic Development, Canada and by the Province of Ontario through the Ministry of Research, Innovation and Science.

\end{document}